\documentclass[aps,prl,twocolumn,longbibliography]{revtex4-1}

\usepackage{amsmath}
\usepackage{mathtools}
\usepackage{braket}
\usepackage{bbm}
\usepackage{graphicx}
\usepackage{amssymb}
\usepackage[dvipsnames]{xcolor}
\usepackage{multirow}
\usepackage{xfrac}
\usepackage{textcomp}

\usepackage{hyperref}
\usepackage{bbold}

\newcommand{\plushalf}{\sfrac{1}{2}}
\newcommand{\minushalf}{{\sfrac{\scalebox{0.75}[1.0]{\( - \)} 1}{2}}}

\hypersetup{
	colorlinks=true,
	linkcolor=blue,
	citecolor=blue,
	filecolor=magenta,
	urlcolor=cyan,
	breaklinks=true
}
\DeclareMathOperator{\Tr}{Tr}

\newcommand{\papertitle}{Fragility of Time-Reversal Symmetry Protected Topological Phases}
\newcommand{\tcm}{T.C.M. Group, Cavendish Laboratory, University of Cambridge, JJ Thomson Avenue, Cambridge, CB3 0HE, U.K.}
\def\authornames{Max McGinley and Nigel R.~Cooper}

\DeclareSymbolFont{sfletters}{OML}{cmbrm}{m}{it}

\newcommand{\iu}{{\rm i}}

\DeclareMathSymbol{\matrrho}{\mathord}{sfletters}{"1A}
\newcommand{\diff}{\mathrm{d}}

\newcounter{firstbib}

\begin{document}
	
	\title{\papertitle}
	\author{Max McGinley}
	\affiliation{\tcm}
	\author{Nigel R. Cooper}
	\affiliation{\tcm}
	
	\date{\today}
	
	\maketitle
	
	\textbf{The second law of thermodynamics points to the existence of an `arrow of time', along which entropy only increases. This arises despite the time-reversal symmetry (TRS) of the microscopic laws of nature. Within quantum theory, TRS underpins many interesting phenomena, most notably topological insulators \cite{Kane2005,Fu2007,Konig2007,Hsieh2008} and the Haldane phase of quantum magnets \cite{Haldane1983a,Pollmann2010}. Here, we demonstrate that such TRS-protected effects are fundamentally unstable against coupling to an environment. Irrespective of the microscopic symmetries, interactions between a quantum system and its surroundings facilitate processes which would be forbidden by TRS in an isolated system. This leads not only to entanglement entropy production and the emergence of macroscopic irreversibility \cite{Neumann1929,Goldstein2010,Srednicki1999}, but also to the demise of TRS-protected phenomena, including those associated with certain symmetry-protected topological phases.  Our results highlight the enigmatic nature of TRS in quantum mechanics, and elucidate potential challenges in utilising topological systems for quantum technologies.
	}
	
	Many isolated systems possess features that rely on symmetries of their Hamiltonian. Most strikingly, in many-body systems the presence of symmetries leads to new phases of matter, including symmetry-protected topological phases (SPTs) \cite{Chen2010,Senthil2015}. SPTs exhibit many remarkable features, such as the emergence of topological bound states (e.g.~Majorana zero modes \cite{Kitaev2001}), which have potential applications in quantum information processing \cite{Alicea2012,Sarma2015}. 
	
	An important practical question, which we address in this Letter, is whether symmetry-protected phenomena such as these can persist in realistic scenarios where the system is weakly coupled to an environment. Previous studies of topology in open systems begin with an approximate equation of motion for the system (e.g.~non-Hermitian Hamiltonian \cite{Bergholtz2019} or Lindblad master equation \cite{Diehl2011,Bardyn2013,Lieu2019}). Instead, our starting point is the full system-environment Hamiltonian
	\begin{align}
	\hat{H}_{\rm tot} = \hat{H}_{S} \otimes \hat{\mathbbm{1}}_{E} + \hat{\mathbbm{1}}_{S} \otimes \hat{H}_{E} + \hat{H}_{SE},
	\label{eqHamTot}
	\end{align}
	where $\hat{H}_S$ and $\hat{H}_{E}$ act on the system and environment, respectively, and $\hat{H}_{SE}$ couples the two. This coupling can always be decomposed as \cite{Breuer2002}
	\begin{align}
	\hat{H}_{SE} = \sum_{\alpha = 1}^M \hat{A}_\alpha \otimes \hat{B}_\alpha,
	\label{eqDecomp}
	\end{align}
	where $\hat{A}_\alpha$, $\hat{B}_\alpha$ are Hermitian operators acting on the system and environment, respectively, and $M$ is the number of `coupling channels'. This approach allows us to define symmetries microscopically, rather than imposing them \textit{a posteriori} on the effective master equation (as in, e.g.~Ref.~\cite{Buca2012}).
	
	Any symmetry-protected features exhibited by $\hat{H}_S$
	will of course be spoiled if $\hat{H}_{SE}$ breaks the relevant symmetries. Even for symmetry-respecting $\hat{H}_{SE}$, the same will still occur if the individual operators $\hat{A}_\alpha$ acting on the system are symmetry-violating.
	For example, tunnelling of electrons in/out of a system can lead to decoherence of Majorana zero modes \cite{Rainis2012}, even when the fermion parity of the combined system and environment is conserved. Therefore, to preserve the desired features, these processes must be suppressed 
	such that the remaining operators $\hat{A}_\alpha$, $\hat{B}_\alpha$, and $\hat{H}_E$ are individually symmetry-respecting. One might expect that scenarios of this type, which we focus on throughout this paper, are sufficiently protected, since each $\hat{A}_\alpha$ obeys the same constraints as the original Hamiltonian $\hat{H}_S$.
	Our key finding is that this intuition can fail: when the protecting symmetry is antiunitary (e.g.~TRS), protection is lost \textit{regardless of the symmetries of} $\hat{A}_\alpha$.
	

	As a concrete example, we focus on the coherence properties of topological bound states. We find that bound states protected by antiunitary symmetries will inevitably decohere at a rate that scales only algebraically with the environment temperature $\tau_{\text{coh}} \sim T^{-\gamma}$ [Eq.~\eqref{eqScaling}] -- this calls into question their potential usefulness in quantum information technologies \cite{Alicea2012,Sarma2015}. In contrast, decoherence processes are thermally activated when the protecting symmetries are unitary $\tau_{\text{coh}} \sim e^{E_{\rm g}/T}$, where $E_{\rm g}$ is the bulk gap. We postulate that corrections to quantized transport in higher dimensional SPTs follow the same pattern of temperature dependence.
	
	To understand this fragility of TRS-protected phenomena, it is instructive to analyse a simple few-body model. Consider an isolated spin-3/2 with Hamiltonian $\hat{H}_S = E_{\rm g}(\hat{S}^z)^2$, with twofold degenerate ground states $\ket{\plushalf}$ and $\ket{\minushalf}$. As long as a suitable symmetry is enforced, the two ground states will remain degenerate when $\hat{H}_S$ is varied. For instance, the degeneracy can be protected by TRS (Kramers' theorem). This eigenstate property is reflected in the dynamics of the system. Consider encoding a qubit in the degenerate subspace, $\ket{\psi}_S = \alpha \ket{\plushalf} + \beta \ket{\minushalf}$. Time evolution under $\hat{H}_S$ leaves this state undisturbed and the qubit can be reliably recovered at late times. Even if $\hat{H}_S$ is weakly perturbed, the overlap $|\braket{\psi(0)|\psi(t)}|^2$ will remain close to 1 provided the appropriate symmetries are maintained.

	How does this change when the spin is weakly coupled to an environment? Insight can be gained from considering the limit $\hat{H}_E = 0$, wherein $\ket{\Psi(t)}$ can be computed using time-dependent perturbation theory in $V \sim \|\hat{H}_{SE}\|$, the characteristic strength of the system-environment coupling. (We will restore $\hat{H}_E$ in a more quantitative calculation later.)
	
	Starting from a factorized initial state $\ket{\Psi(0)} = \ket{\psi}_S \otimes \ket{\chi}_E$, the correction to first order in $V$ is $\ket{\Psi^{(1)}(t)} = -\iu t \sum_{\alpha} \hat{\Pi}_{\rm GS} \hat{A}_\alpha \ket{\psi}_S \otimes \hat{B}_\alpha \ket{\phi}_E$, where $\hat{\Pi}_{\rm GS}$ projects onto the degenerate ground state subspace of the system $S$. For generic $\{\hat{A}_\alpha\}$, the system becomes entangled with the environment (since $\ket{\Psi(t)}$ cannot be written in a factorized form), leading to decoherence of the qubit. Note that decoherence still occurs even if $\hat{H}_{SE}$ is itself symmetric. However, if all $\{\hat{A}_\alpha\}$ respect the same symmetry as the Hamiltonian $\hat{H}_S$, then these operators can only act trivially within the degenerate subspace, i.e.~$\hat{\Pi}_{\rm GS} \hat{A}_\alpha \ket{\psi}_S = a_\alpha \ket{\psi}_S$ \cite{Yang2017}. This gives $\ket{\Psi(t)} = \ket{\psi}_S \otimes (1 - \iu t \sum_\alpha a_\alpha \hat{B}_\alpha)\ket{\phi}_E$, so the system remains unperturbed. This lends credence to the simple expectation, stated above, that coherence is preserved if the operators  $\{\hat{A}_\alpha\}$ are invariant under the symmetries of $\hat{H}_{\rm S}$ that protect the degeneracy.
	
\begin{figure}
		\includegraphics[scale=1]{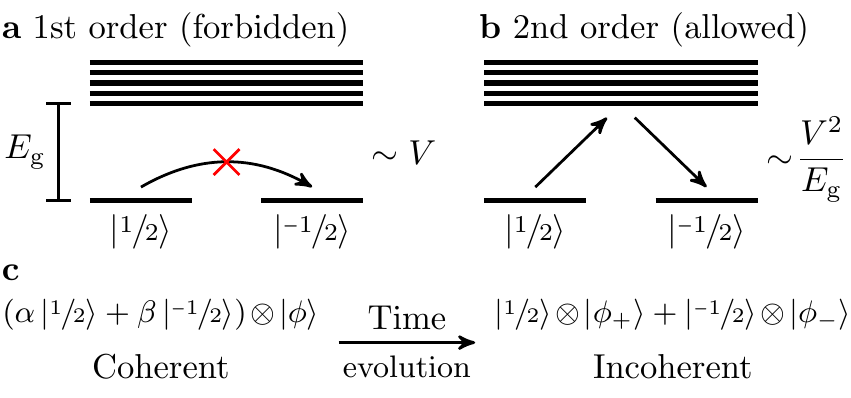}
		\caption{\textbf{Decoherence mechanisms for topological bound states coupled to an environment.} The spectrum of the 1D SPT system $\hat{H}_S$ (thick lines) features degenerate ground states ($\ket{\plushalf}$ and $\ket{\protect\minushalf}$, representing different configurations of the bound state), which are separated in energy from bulk excitations by a gap $E_{\rm g}$. If the environment is at a temperature $T \ll E_{\rm g}$, then transitions to excited states are thermally activated, and occur at an exponentially slow rate $\sim e^{-E_{\rm g}/T}$. \textbf{a}, Direct transitions between ground states are forbidden by symmetry if the coupling operators $\hat{A}_\alpha$ [Eq.~\eqref{eqDecomp}] respect the relevant symmetries. \textbf{b}, If the protecting symmetry is TRS (or any antiunitary symmetry), then indirect transitions are allowed regardless of the symmetries of $\hat{A}_\alpha$. These proceed via a virtual excited state, and the corresponding matrix elements scale as $V^2/E_{\rm g}$ [see Eq.~\eqref{eqSecond}]. \textbf{c}, Because TRS acts non-trivially on both the system and environment, the coherence of the bound state (being a property of the system only) is no longer protected and the initial qubit decoheres. This obstruction to defining a `local' TRS, acting on a subspace of the total Hilbert space, can be used to understand how an arrow of time emerges from TRS-respecting laws of motion \cite{Srednicki1999,Neumann1929,Goldstein2010}.}
		\label{figTransition}
	\end{figure}

	However, this hypothesis turns out to be incorrect in general. This can be seen already from the second order corrections in $V$:
	\begin{align}
	\ket{\Psi^{(2)}(t)} &= \frac{-\iu t}{E_{\rm g}} \sum_{\alpha \beta} \hat{\Pi}_{\rm GS} \hat{A}_{\alpha} \hat{\Pi}_{\rm Ex} \hat{A}_{\beta} \ket{\psi}_S\otimes \hat{B}_{\alpha} \hat{B}_{\beta} \ket{\phi}_E,
	\label{eqSecond}
	\end{align}
	where $\hat{\Pi}_{\rm Ex} \coloneqq \hat{\mathbb{1}} - \hat{\Pi}_{\rm GS}$ projects onto excited states. (We have assumed that the coupling is gradually turned on at a rate slower than $E_{\rm g}$, and ignored contributions $\propto \ket{\Psi(0)}$.) Equation \eqref{eqSecond} captures processes that occur via a virtual excited state [see Fig.~\ref{figTransition}b].
	
	By analogy to the above, transitions will only occur if $\hat{C}_{\alpha \beta} \coloneqq \hat{A}_{\alpha} \hat{\Pi}_{\rm Ex} \hat{A}_{\beta}$ acts non-trivially within the ground state subspace. Observe that $\hat{C}_{\alpha \beta}$ is itself invariant under the relevant symmetries; however, it is generically non-Hermitian, and so might not obey the same constraints as a symmetry-respecting Hamiltonian. We therefore decompose $\hat{C}_{\alpha \beta} = \hat{X}_{\alpha \beta} + \iu \hat{Y}_{\alpha \beta}$, where $\hat{X}_{\alpha \beta} \coloneqq (\hat{C}_{\alpha \beta} + \hat{C}_{\beta \alpha})/2$, and $\hat{Y}_{\alpha \beta} \coloneqq -\iu (\hat{C}_{\alpha \beta} - \hat{C}_{\beta \alpha})/2$ are both Hermitian. Now, if the protecting symmetries are unitary, then both $\hat{X}_{\alpha \beta}$ and $\hat{Y}_{\alpha \beta}$ are also symmetry-respecting Hermitian operators, and so cannot cause transitions between different ground states. We show in the Methods section that transitions among ground states are forbidden at all orders in $V$, and so the system and environment remain unentangled. However, due to the factor of $(-\iu)$ required by Hermiticity, $\hat{Y}_{\alpha \beta}$ will \textit{not} be invariant under antiunitary symmetries, such as time-reversal. If the ground state degeneracy is protected by antiunitary symmetries, then $\hat{C}_{\alpha \beta}$ \textit{can} act non-trivially within the ground state subspace for $\alpha \neq \beta$. (For example, take $\hat{A}_1 = (\hat{S}^x)^2$ and $\hat{A}_2 = \{\hat{S}^x, \hat{S}^z\}$, which are both TRS-even.) Unless $\hat{H}_{SE}$ is fine-tuned such that it is factorizable, i.e.~$M = 1$, then this leads to decoherence of the qubit. Although the limit $\hat{H}_E = 0$ precludes an estimation of a corresponding decoherence rate, we see that the perfect coherence enjoyed by the isolated system is fragile against coupling to an environment if the protecting symmetries are antiunitary. This decoherence is also manifest in the eigenstates of $\hat{H}_{\rm tot}$; see Methods.
	
	While the above analysis refers explicitly to the spin-3/2 model, it highlights a much more general issue regarding symmetry protection in quantum systems. The problem stems from the fact that there is no way to consistently define antiunitary symmetries on a subsystem of some larger Hilbert space (see, e.g.~Ref.~\cite{Pollmann2012a}, p.~8). Consequently, the system-environment coupling will enable processes that effectively break TRS for the system, regardless of any microscopic symmetry considerations. The r{\^o}le of this mechanism in the emergence of irreversible dynamics is well-known. Here, we show that it leads to an inherent fragility of TRS-protected phenomena:
	In the above, even if every component of the Hamiltonian ($\{\hat{A}_\alpha\}$, $\{\hat{B}_\alpha\}$, $\hat{H}_S$, and $\hat{H}_E$) were TRS-invariant, the relevant protection occurs not at the level of the system Hilbert space, but on the composite system-environment Hilbert space. Thus, without explicit control over the environment, the system will not exhibit the desired TRS-protected properties (e.g.~coherence of quantum information, see Fig.~\ref{figTransition}c). In contrast, it \textit{is} possible to define a unitary symmetry that pertains only to the system and not to the environment, under which the relevant phenomena can remain protected at non-zero coupling.

	Our arguments are readily extended to symmetry-protected topological phases (SPTs). In isolated one-dimensional SPTs, the system boundaries host topological bound states -- collective degrees of freedom that remain spatially localized and gapless as long as the relevant symmetries are enforced.
	We will focus on dynamics in the vicinity of one such bound state; accordingly, the eigenstate structure of the system exactly mirrors that of the spin-3/2: There are multiple ground states (representing different configurations of the bound state) whose degeneracy is protected by a group of symmetries, and all excited states have energies above some gap $E_{\rm g}$ (see Fig.~\ref{figTransition}). Our newfound intuition suggests that if the SPT is protected by (anti-)unitary symmetries, then the topological bound state will (not) remain coherent upon coupling to an environment. More precisely, those phases that can be trivialized by explicitly breaking all antiunitary symmetries will exhibit decoherence; this particular class of SPTs has been classified in a different context \cite{McGinley2019,McGinley2019a}.
	
	We confirm this by explicitly calculating a decoherence rate for quantum information stored within the bound state. Here, we no longer neglect $\hat{H}_E$ (which itself will be symmetry-respecting), and consider a thermal environment $\hat{\rho}(0) = \ket{\psi}_S\bra{\psi}_S \otimes \hat{\rho}_E$, where $\hat{\rho}_E \propto e^{-\hat{H}_E/T}$. Moreover, we focus on the regime $T \ll E_{\rm g}$, such that transitions to excited states are exponentially slow $\tau_{\text{ex}} \sim e^{E_{\rm g}/T}$. (The effects of thermally generated excitations on bound state coherence have been considered elsewhere \cite{Goldstein2011}.) Our calculation, described in the methods section, resembles that of the spin-3/2 in terms of symmetry considerations. However, rather than computing $\ket{\Psi(t)}$, we derive a master equation for the system density matrix $\hat{\rho}_S(t) \coloneqq \Tr_E \hat{\rho}(t)$. As before, we must account for transitions between ground states proceeding via a virtual excited state. We therefore work beyond the commonly employed Born-Markov approximation \cite{Breuer2002}, which captures only lowest-order effects. For a bound state protected by antiunitary symmetries coupled to the simplest type of environment (a bath of harmonic oscillators), we find that at leading order in $V$, $\tau_{\text{coh}}$ scales as
	\begin{align}
	\tau_\text{coh} \sim \frac{E_{\rm g}^4 \omega_\text{c}^{2+2s}}{V^4T^{3+2s}\vphantom{^\dagger}},
	\label{eqScaling}
	\end{align}
	where the exponent $s$ and cutoff frequency $\omega_\text{c}$ characterise the distribution of oscillator frequencies in the bath (see Methods). Although the exact dependence on $T$ may vary slightly for more structured environments, crucially it is only algebraic. In contrast, when the protecting symmetry is unitary, the fastest decoherence process involves propagation of a bulk excitation across the system \cite{Rainis2012}; this thermally activated processes is exponentially slow $\tau_{\text{coh}} \sim e^{E_{\rm g}/T}$. As well as dictating the lifetime of quantum information, $\tau_{\text{coh}}^{-1}$ could also be inferred from spectroscopic measurements of the system as a characteristic width of the zero-energy peak \cite{Lieu2019}. 
	
	Higher dimensional SPTs possess gapless edge modes which give rise to quantized transport signatures. For example, in an isolated quantum spin Hall bar, TRS forbids elastic backscattering between counter-propagating edge channels, leading to perfect conduction \cite{Kane2005}. It is well known that quantization can be marred by environmental couplings for which $\hat{A}_\alpha$ are symmetry-breaking (e.g.~magnetic impurities \cite{Maciejko2009} or tunnelling into leads \cite{Roth2009}); however our findings demonstrate that such TRS-breaking processes occur much more ubiquitously. While an explicit conductance calculation is beyond the scope of this work, our arguments can be used to show that \textit{elastic} backscattering between degenerate counter-propagating states in helical channels can occur via the same virtual transition that led to decoherence of topological bound states in the above, even for a bosonic, non-magnetic environment. (Note that this effect differs in nature from the \textit{inelastic} backscattering processes which have previously been identified \cite{Schmidt2012a,Budich2012a,Vayrynen2018}.) The quantized conductivity protected by TRS is thus in this sense fragile against coupling to an environment. Conductance quantization would be restored if an appropriate unitary symmetry were additionally imposed (e.g.~if spin orbit coupling vanishes so that total spin is conserved).
	
	In conclusion, we have argued on general grounds that phenomena protected by TRS (or other antiunitary symmetries) are inevitably compromised by coupling to an environment. We attributed this effect to the fact that such symmetries cannot be defined on a subsystem of a larger Hilbert space. Thus, although a given composite system may respect TRS microscopically, any subsystem therein can propagate in a seemingly TRS-violating manner, since it is not itself isolated. This leads to both the emergence of macroscopic irreversibility, and to the inevitable loss of TRS-protected phenomena in open systems.\\ 
	
	\textbf{Acknowledgements.} This work was supported by an EPSRC studentship and Grants No.~EP/P034616/1 and No.~EP/P009565/1, and by an Investigator Award of the Simons Foundation.
	
	\textbf{Author Contributions.} Both authors contributed to the formulation of the study, the interpretation of the results and the writing of the manuscript. M.M.~developed and performed the calculations.
	
	\textbf{Data availability.} Data sharing is not applicable to this article as no datasets were generated or analysed during the current study.

	\newpage
	\appendix

%
%
	
	\onecolumngrid
	
	\begin{center}
		{\fontsize{12}{12}\selectfont
			\textbf{Methods for ``{\papertitle}''\\[5mm]}}
		{\normalsize \authornames\\[1mm]}
		{\fontsize{9}{9}\selectfont  
			\textit{\tcm}}
	\end{center}
	\normalsize
	
	\twocolumngrid
	
	\subsection*{Coherence Time of Topological Bound States}
	
	Here, we outline the calculation of the coherence time for a topological bound state weakly coupled to an environment in thermal equilibrium at a temperature $T \ll E_{\rm g}$. The calculation in the main text elucidates the structure of matrix elements between states of the system due to the coupling $\hat{H}_{SE}$. However, there we took a simplifying limit $\hat{H}_E = 0$, which led to an unusual time-dependence of the transition probabilities ($P_{i \rightarrow f}(t) \propto t^2$, rather than the familiar Fermi's Golden rule result $P_{i \rightarrow f}(t) = \gamma t$). Here, we will include $\hat{H}_E$, which will lead to well-defined transition rates $t^{-1}P_{i \rightarrow f}(t)$; however our key findings regarding the differences between unitary and antiunitary symmetries do not change.
	
	For concreteness, let us describe the symmetry properties of a topological bound state. The Hamiltonian $\hat{H}_S$ will possess $N_S$ ground states $\hat{H}_{S}\ket{j} = 0$, $j = 1,\ldots, N_{S}$, each differing only in the vicinity of the bound state under consideration.  The ground state subspace $\mathcal{H}_{\text{GS}} = \text{span}(\ket{j})$ forms a $N_S$-dimensional irreducible projective representation of the protecting symmetry group $G$. Accordingly, any symmetry-respecting Hermitian operator $\hat{H}$ must satisfy $\hat{\Pi}_{\rm GS} \hat{H}\, \hat{\Pi}_{\rm GS} \propto \hat{\Pi}_{\rm GS}$ (this is a consequence of Schur's lemma \cite{Yang2017}). The same structures arise in systems possessing Majorana zero modes, although one may need to keep track of an additional bound state far from the region of interest, such that the system is composed of a whole number of Dirac fermions.
	
	In our calculation for the open system, we will make use of the two-time correlation functions
	\begin{align}
	\tilde{\Gamma}_{\alpha \beta}(t) \coloneqq \Tr_E\left(\hat{\rho}_E \hat{B}_\alpha(t) \hat{B}_\beta(0) \right) = \int \frac{\diff \epsilon}{2\pi} e^{-\iu \epsilon t} \Gamma_{\alpha \beta}(\epsilon),
	\label{eqCorrelator}
	\end{align}
	and the associated spectral functions $\Gamma_{\alpha \beta}(\epsilon)$. (Here, $\hat{B}_\alpha(t) \coloneqq e^{\iu \hat{H}_E t} \hat{B}_\alpha e^{-\iu \hat{H}_E t}$.) For simplicity we assume that the environment is Gaussian, such that the above quantities fully characterise the state of the environment $\hat{\rho}_E = e^{-\beta \hat{H}_E}/Z$, where $Z = \Tr_E e^{-\beta \hat{H}_E}$ is the partition function. 
	Spectral functions will be exponentially suppressed for large negative arguments $\Gamma_{\alpha \beta}(-|\epsilon|) \sim e^{-\beta|\epsilon|}$. We will therefore neglect contributions to the decoherence rate for which $\Gamma_{\alpha \beta}(\epsilon)$ is evaluated at $\epsilon \leq -E_{\rm g}$ (these terms will represent the generation of bulk excitations
	).
	
	Our aim will be to derive a master equation for the state of the system $\hat{\rho}_S(t) = \Tr_E \hat{\rho}(t)$. In the scenarios considered in this Letter, the dynamics of $\hat{\rho}_S(t)$ is well described by a quantum Markov process  over appropriately coarse-grained timescales; that is $\partial_t \hat{\rho}(t) \approx \mathcal{L} \hat{\rho}(t)$, where the time-independent generator $\mathcal{L}$ is an appropriate superoperator. To understand why this is so, we must compare the typical rate of change of $\hat{\rho}_S(t)$ (given by $\tau_{\text{coh}}^{-1}$) with the `memory time' of the environment $\tau_\text{m}$, i.e.~the characteristic timescale over which $\Gamma_{\alpha \beta}(t)$ decays. If $\tau_{\text{coh}} \gg \tau_\text{m}$ [which does indeed turn out to be true, as can be seen from \eqref{eqScaling}], then the back-action of the system on the environment is `forgotten' before the system has changed appreciably. Accordingly, provided one is not interested in the temporal variation of $\hat{\rho}_S(t)$ over timescales shorter than $\tau_{\rm m}$, the dynamics of the system can be assumed to be independent of its history. (See Ref.~\cite{Kampen1974} for a fuller discussion of an analogous classical problem.)
	
	With this understood, we can calculate the generator $\mathcal{L}$ by calculating $\hat{\rho}_S(t)$ (coarse grained over a timescale $\Delta t \gg \tau_{\rm m}$), and comparing it with the formal solution $\hat{\rho}(t) = e^{\mathcal{L} t} \hat{\rho}(0)$. Specifically, for $t \ll \tau_{\text{coh}}$, we expect that the linear-in-time component of $\hat{\rho}(t)$ will be exactly $t \times \mathcal{L} \hat{\rho}(0)$. (This is analogous to the derivation of transition rates in Fermi's Golden Rule.) We will find that $\tau_{\text{coh}} \gg V^{-1}$ (where $V \sim \| \hat{H}_{SE}\|$ is the system-environment coupling strength), and so on these timescales, we expect that time-dependent perturbation theory will converge well. We will work in the interaction picture with respect to $\hat{H}_0 = \hat{H}_S + \hat{H}_E$, such that $\hat{\rho}(t) = \hat{U}(t,0) \hat{\rho}(0) \hat{U}^\dagger(t,0)$, where the time evolution operator is $\hat{U}(t,t') = \mathcal{T} \exp[-\iu \int_{t'}^t \diff t_1 \hat{H}_I(t_1)]$ (where $\mathcal{T}$ denotes time-ordering, and $\hat{H}_I(t) = e^{\iu \hat{H}_0 t}\hat{H}_{SE} e^{-\iu \hat{H}_0 t}$).
	
	To proceed, we expand the time-evolution operators either side of $\hat{\rho}(0)$ in the expression for $\hat{\rho}_S(t)$ in powers of $V$, and then take the trace over environment degrees of freedom. The derivation of the lowest order ($V^2$) contributions is well-known \cite{Breuer2002}, and gives
	\begin{align}
	\frac{\diff\hat{\rho}_S}{\diff t}  &= \sum_{\omega, \alpha, \beta} \Gamma_{\alpha \beta}(\omega)\left[ \hat{A}_\beta(\omega) \hat{\rho} \hat{A}_\alpha^\dagger (\omega)
	\right.\nonumber\\ &-\left.  \frac{1}{2}\left\{ \hat{A}_\alpha^\dagger(\omega) \hat{A}_\beta(\omega), \hat{\rho}_S  \right\} \right]
	+ \iu[\hat{H}_S + \hat{H}_{LS},\hat{\rho}_S].
	\label{eqLowest}
	\end{align}
	Here, $\{\cdot, \cdot\}$ is the anticommutator, and  $\hat{A}_\alpha(\omega) = \sum_{\epsilon' - \epsilon = \omega} \hat{\Pi}_\epsilon \hat{A}_\alpha \hat{\Pi}_{\epsilon'}$ is the component of $\hat{A}_\alpha$ that lowers the energy of the system by an amount $\omega$ \cite{Breuer2002} ($\hat{\Pi}_\epsilon$ is the projector onto the eigenspace of $\hat{H}_S$ with energy $\epsilon$). For our purposes, we need only know that the Lamb shift $\hat{H}_{LS}$ is Hermitian; commutes with $\hat{H}_S$; and respects the same symmetries as $\hat{H}_{SE}$. Therefore it will have no effect within the ground state subspace. The remaining part of Eq.~\eqref{eqLowest} captures direct processes in which an energy $\omega$ is transferred from system to environment; thus if the initial state is a ground state, the only non-thermally activated processes will be those with $\omega = 0$, and transitions will be generated by $\hat{A}_\alpha(0) \hat{\Pi}_{GS} = \hat{\Pi}_{GS} \hat{A}_\alpha \hat{\Pi}_{GS}$, as we found for the simpler model in the main text. If $\{\hat{A}_\alpha\}$ respect the relevant symmetries, then the above gives $\diff \hat{\rho}_S / \diff t = 0$, up to corrections that scale as $e^{-E_{\rm g}/T}$.
	
	The intuition developed in the main texts suggests that when the protecting symmetries are antiunitary, decoherence will arise at next-to-leading order, which requires us to calculate $\hat{\rho}(t)$ to fourth order in $V$. If we write $\hat{\rho}^{(i,j)}_S(t)$ for the contribution coming from expanding $\hat{U}(t,0)$ to $i$th order and $\hat{U}^\dagger(t,0)$ to $j$th order, then one of the contributing terms is $\hat{\rho}^{(2,2)}_S(t)$:
	\begin{align}
	\hat{\rho}^{(2,2)}_S(t) &=\sum_{\alpha_1 \ldots \alpha_4} \sum_{\omega_1 \ldots \omega_4} \int_{0}^t \diff t_1\, \int_{0}^{t_1} \diff t_2\, \int_{0}^t \diff t_3\, \int_{0}^{t_3} \diff t_4 \nonumber\\
	&\times \hat{A}_{\alpha_1}(\omega_1) \hat{A}_{\alpha_2}(\omega_2)\ket{\psi_S(0)} \bra{\psi_S(0)} \hat{A}_{\alpha_4}^\dagger(\omega_4) \hat{A}_{\alpha_3}^\dagger(\omega_3) \nonumber\\ &\times \Tr_E[ \hat{B}^\dagger_{\alpha_4}(t_4) \hat{B}^\dagger_{\alpha_3}(t_3) \hat{B}_{\alpha_1}(t_1) \hat{B}_{\alpha_2}(t_2) \hat{\rho}_E] \nonumber\\
	&\times e^{-\iu \omega_1 t_1 - \iu \omega_2 t_2 + \iu \omega_3 t_3 + \iu \omega_4 t_4}.
	\label{eqContrib}
	\end{align}
	The trace over the environment can be expressed in terms of the correlation functions $\tilde{\Gamma}_{\alpha \beta}(t)$ by using our assumption that $\hat{\rho}_E$ and $\hat{H}_E$ are Gaussian.
	
	In our setup, the initial state of the system $\ket{\psi_S(0)}$ is a ground state of $\hat{H}_S$, which is separated in energy from excited states by a gap $E_{\rm g}$. For each term in the sum over $\{\omega_i\}$, we therefore have either $\omega_1 + \omega_2 = 0$, or $\omega_1 + \omega_2 \leq -E_{\rm g}$ (similarly for $\omega_3 + \omega_4$). The latter terms, which correspond to bulk excitations, can be neglected, since to make such a term on-shell requires the environment to provide an energy $E_{\rm g} \gg T$, which will be suppressed as $\Gamma_{\alpha \beta}(-E_{\rm g}) \sim e^{-E_{\rm g}/T}$. (This still leaves off-shell contributions, but these should not be included when coarse-graining over a timescale $\Delta t \gg \tau_{\rm m}$, since they oscillate at a rate much faster than $\tau_{\rm m}^{-1}$. This coarse-graining can be performed by considering the Laplace transform of \eqref{eqContrib} at values of the Laplace parameter much less than $(\Delta t)^{-1}$.) After a lengthy yet straightforward derivation, including the other $\hat{\rho}_S^{(i,j)}(t)$, and using the realness of $\Gamma_{\alpha \beta}(\epsilon)$ (which follows from the time-reversal symmetry of $\hat{H}_E$), we arrive at an expression for $\hat{\rho}_S(t)$ from which we infer that the master equation is
	\begin{widetext}
		\begin{align}
		\frac{\diff \hat{\rho}_S}{\diff t} &= \sum_{\{\alpha_i\}} \sum_{\omega_1, \omega_2 \geq E_{\rm g}}  \int \frac{\diff \epsilon}{4\pi} \Gamma_{\alpha_4 \alpha_1}(\epsilon) \Gamma_{\alpha_3 \alpha_2}(-\epsilon)
		\left(\hat{C}_{\alpha_1 \alpha_2}(\omega_1,\epsilon)\hat{\rho}_S\, \hat{C}_{\alpha_4 \alpha_3}(\omega_2,-\epsilon) - \frac{1}{2}\left\{ \hat{C}_{\alpha_1 \alpha_2}(\omega_1,\epsilon) \hat{C}_{\alpha_4 \alpha_3}(\omega_2,-\epsilon)\, , \, \hat{\rho}_S \right\} \right),
		\label{eqDissipation}
		\end{align}
	\end{widetext}
	where
	\begin{align}
	\hat{C}_{\alpha \beta}(\omega,\epsilon) \coloneqq \hat{\Pi}_{\rm GS}\left[ \frac{\hat{A}_{\alpha}(\omega)  \hat{A}_{\beta}^\dagger(\omega)}{\omega - \epsilon} + \frac{\hat{A}_{\beta}(\omega)  \hat{A}_{\alpha}^\dagger(\omega)}{\omega + \epsilon} \right] \hat{\Pi}_{\rm GS}.
	\label{eqCommutator}
	\end{align}
	The quantity \eqref{eqCommutator} generalises the operator $\hat{C}_{\alpha \beta}$ which we defined in the main text. Again, if the symmetries protecting the topological bound state are unitary, then both the Hermitian and antihermitian components of $\hat{C}_{\alpha \beta}(\omega, \epsilon)$ are constrained by Schur's Lemma, and so $\hat{C}_{\alpha \beta}(\omega, \epsilon) \propto \hat{\Pi}_{\rm GS}$; in this case the above reduces to $\diff \hat{\rho}_S/\diff t = 0$, and we conclude that the bound state can only decohere through thermally activated processes at this order. Moreover, if we were to compute the master equation at $(2n)$th order in $V$, we see from the structure of perturbation theory that non-thermally-activated transitions would be generated by analogous operators $\hat{C}_{\alpha_1 \ldots \alpha_n}$ composed of products of $n$ operators $\hat{\Pi}_{\rm GS} \hat{A}_{\alpha_1}(\omega_1) \cdots \hat{A}_{\alpha_n}(\omega_n) \hat{\Pi}_{\rm GS}$, which are projected onto the ground state subspace to ensure conservation of energy. Any such product can be decomposed into Hermitian and antihermitian components, which again must both be proportional to $\hat{\Pi}_{\rm GS}$ by Schur's Lemma, and thus will be unable to cause transitions. We conclude that for unitary symmetries, the coherence time scales as $\tau_{\text{coh}} \sim e^{E_{\rm g}/T}$ at all orders in perturbation theory.
	
	In contrast, if an antiunitary symmetry is required to protect the bound state, then the antihermitian component $\hat{Y}_{\alpha \beta}(\omega,\epsilon) \coloneqq -\iu[\hat{C}_{\alpha \beta}(\omega,\epsilon) - \hat{C}_{\beta \alpha}(\omega,\epsilon)]/2$ can act non-trivially within the ground state subspace. In this case, $\tau_{\text{coh}}$ is not thermally activated. The integral in \eqref{eqDissipation} will be dominated by the region $|\epsilon| \lesssim T \ll E_{\rm g}$, and so we can expand the energy denominators appearing in \eqref{eqCommutator} in powers of $\epsilon/\omega$. The zeroth order terms are Hermitian, and so do not contribute to $\hat{Y}_{\alpha \beta}(\omega,\epsilon)$. We therefore have $\hat{Y}_{\alpha \beta}(\omega,\epsilon) \approx (\epsilon/\omega^2) \hat{D}_{\alpha \beta}(\omega)$ for some appropriate $\epsilon$-independent dimensionless operator $\hat{D}_{\alpha \beta}(\omega)$, up to corrections that are higher order in $T/E_{\rm g}$. Since $\omega \gtrsim E_{\rm g}$, the decoherence rate is on the same order as the integral $K_{\{\alpha_i\}} \coloneqq E_{\rm g}^{-4} \int \diff \epsilon\, \epsilon^2 \Gamma_{\alpha_4 \alpha_1}(\epsilon) \Gamma_{\alpha_3 \alpha_2}(-\epsilon) $.
	
	We can estimate $K_{\{\alpha_i\}}$ in the case where the environment is a bath of harmonic oscillators $\hat{H}_E = \sum_q \omega_q \hat{b}_q^\dagger \hat{b}_q$ (with canonical commutation relations $[\hat{b}_q, \hat{b}_{q'}^\dagger] = \delta_{qq'}$), and the couplings are linear $\hat{B}_\alpha = \sum_q g_{\alpha q} \hat{b}_q + g_{\alpha q}^* \hat{b}_q^\dagger$. Following Caldeira and Leggett \cite{Caldeira1981}, we define the bath spectral density $J_{\alpha \beta}(\omega) \coloneqq \sum_q g_{\alpha q}^* g_{\beta q} \delta(\omega - \omega_q)$. The spectral functions are then given by $\Gamma_{\alpha \beta}(\omega) = \Theta(\omega) [1+n_B(\omega)]J_{\alpha \beta}(\omega) + \Theta(-\omega) n_B(-\omega)J_{\beta \alpha}(-\omega)$, where $n_B(\omega) = (e^{\omega/T} - 1)^{-1}$ is the Bose distribution function. The bath spectral density is normalised such that $\int_0^\infty \diff \omega J_{\alpha \beta}(\omega) = \Tr[\hat{B}_\alpha^\dagger \hat{B}_\beta] \sim V^2$, and is typically characterised by a power-law at small frequencies with exponent $s$, and a cutoff at large frequencies $\omega \gtrsim \omega_\text{c}$, e.g.~$J_{\alpha \beta}(\omega) \sim V^2 \omega^s \omega_\text{c}^{-s-1} e^{-\omega/\omega_\text{c}}$ (however only the low-frequency behaviour of $J_{\alpha \beta}(\omega)$ matters here, provided $\omega_\text{c} \gg T$). The case $s = 1$ corresponds to an Ohmic bath. It follows straightforwardly that $K_{\{\alpha_i\}} \approx \kappa_{\{\alpha_i\}} T^{3+2s}V^4 E_{\rm g}^{-4} \omega_\text{c}^{-2-2s}$, where $\kappa_{\{\alpha_i\}}$ are non-universal dimensionless constants of order 1. This justifies the scaling of $\tau_{\text{coh}}$ quoted in Eq.~\eqref{eqScaling}.
	
	Having established the scaling behaviour of the coherence times of topological bound states in general, we provide some specific examples. For TRS-broken topological superconductors possessing Majorana zero modes, the protecting symmetry is conservation of fermion parity, which is unitary. (In a non-interacting system, one may instead view the phase as being protected by particle-hole symmetry, which imposes a constraint on the first quantized Hamiltonian that involves complex conjugation. However as a many-body operator, particle-hole symmetry is still unitary \cite{Chiu2016}, and so should still be robust; see also Ref.~\cite{McGinley2019}.) If fermions can tunnel between system and environment then $\hat{A}_\alpha$ will break the symmetry, and the decoherence rate will be governed by direct processes [Eq.~\eqref{eqLowest}]. This `quasiparticle poisoning' effect \cite{Rainis2012} leads to a finite coherence time $\tau_{\text{coh}} \sim V^2 \Gamma(\omega = 0)$. For example, coupling to metallic leads gives $\tau_{\text{coh}} \sim V^2 \nu(E_F)$, where $\nu(E_F)$ is the density of states at the Fermi energy. If tunnelling is suppressed, e.g.~via Coulomb blockade effects \cite{Oreg2020}, then $\hat{A}_\alpha$ conserve fermion parity, and the coherence time is thermally activated $\tau_{\rm coh} \sim e^{E_{\rm g}/T}$.
	
	Examples of topological bound states protected by antiunitary symmetries include Majorana Kramers' pairs in time-reversal symmetric topological superconductors \cite{Wong2012}; spin-1/2 edge modes in the Haldane phase \cite{Haldane1983a}; and boundary modes of the Su-Schreiffer-Heeger chain \cite{Su1979} (protected by antiunitary chiral symmetry). If no additional symmetries are present, then these modes will possess a non-thermally-activated coherence time; however there may be scenarios where an additional unitary symmetry is present which is sufficient to protect the phase in question, e.g.~spin rotation symmetry can protect the Haldane phase \cite{Pollmann2010}. When the system in question features a combination of unitary and antiunitary symmetries, one can determine whether the coherence will remain protected by consulting the classification tables in Refs.~\cite{McGinley2019,McGinley2019a}, which enumerate those phases that are stable once antiunitary symmetries are removed.
	
	\subsection*{Eigenstates of the Composite System}
	
	As mentioned in the main text, our findings can be understood in a time-independent framework based on eigenstates of the full Hamiltonian \eqref{eqHamTot}. For example, consider the open spin-3/2 model, protected by TRS. We assume that the environment is ergodic, so eigenstates of $\hat{H}_E$ are thermal in the sense of the eigenstate thermalisation hypothesis, and can be assigned a corresponding temperature $T^{-1} = \diff S(E)/\diff T$, where $S(E)$ is the thermodynamic entropy at energy $E$ \cite{Srednicki1999}. (Even if $\hat{H}_E$ were not ergodic, as in the calculation above, we expect that the weak coupling $\hat{H}_{SE}$ will induce ergodicity without changing $S(E)$ appreciably.) When the system-environment coupling is turned on, a given factorized eigenstate of the decoupled system will strongly hybridize with other eigenstates that are nearby in energy. Specifically, we expect strong hybridization when the matrix element coupling the two states is greater than the level spacing in the environment $\delta_E \sim e^{-\text{const} \times N}$ (see Ref.~\cite{Roeck2017} for a related problem). We have seen already that when the protecting symmetry is antiunitary, different ground states can be coupled via indirect processes, with matrix elements of order $V^2/E_{\rm g}$ [Fig.~\ref{figTransition}b]. Therefore, the number of these resonant states contributing to a given eigenstate is $\mathcal{G} \coloneqq V^2 E_{\rm g}^{-1} \delta_E^{-1} \gg 1$. 
	
	Provided $V \ll E_{\rm g}$, a sufficiently low-energy eigenstate can be written as $\ket{\Psi} = \ket{\plushalf} \otimes \ket{\phi_+} + \ket{\minushalf} \otimes \ket{\phi_-}$. (Components of $\ket{\Psi}$ in which the system is excited will be exponentially suppressed $\sim e^{-E_{\rm g}/T}$, provided that the eigenenergy in question corresponds to an environment temperature $T \ll E_{\rm g}$.) Now, since $\hat{H}_{\rm tot}$ is itself TRS-invariant, Kramers theorem can be applied to the composite system and environment, so the eigenstates come in degenerate pairs. However since the protecting symmetry is antiunitary, the operation that relates degenerate eigenstates involves nontrivial transformations on both the system and environment. Therefore, there are no separate symmetry constraints on $\ket{\phi_{\pm}}$. Assuming that the $\sim \mathcal{G}$ unperturbed eigenstates from which $\ket{\Psi}$ is composed are `typical' (i.e.~not fine-tuned), we expect $|\braket{\phi_+|\phi_-}|^2 \sim \mathcal{G}^{-1}$. Therefore, for $\mathcal{G} \gg 1$ the eigenstates of $\hat{H}_{tot}$ will be incoherent mixtures of the two ground states.  Given that the eigenstates dictate the state of the open system at late times, this is consistent with our findings. We can also see that the critical coupling strength where the eigenstates cross over from coherent to incoherent is $V_c \sim \sqrt{E_{\rm g} \delta_E}$, which is exponentially small in the number of degrees of freedom in the environment. Thus an arbitrarily weak system-environment interaction will lead to decoherence of the system in question, provided that the environment is sufficiently large.
	
	In contrast, if an appropriate unitary symmetry were imposed that acts only on the system, then one can readily show that $\ket{\phi_+} = \alpha \ket{\phi_-}$ for some constant $\alpha$, since the symmetry operation leaves the environment unaffected. The eigenstates therefore have vanishing system-environment entanglement, and coherence is maintained.

%
	
	%
	
\end{document}